\documentclass[aps,showpacs]{revtex4}
\usepackage{graphicx}

\input{tcilatex}

\begin{document}

\title{Negative lateral conductivity of hot electrons in a biased
superlattice}
\author{A. Hern\'{a}ndez-Cabrera}
\email{ajhernan@ull.es}
\author{P. Aceituno}
\affiliation{Dpto. F\'{\i}sica B\'{a}sica, Universidad de La Laguna, La Laguna,
38206-Tenerife, Spain}
\author{F.T. Vasko}
\email{ftvasko@yahoo.com}
\affiliation{Institute of Semiconductor Physics, NAS Ukraine, Pr. Nauki 41, Kiev, 03028,
Ukraine}
\date{\today }

\begin{abstract}
Nonequilibrium electron distribution in a superlattice subjected to a
homogeneous electric field (biased superlattice with equipopulated levels)
is studied within the tight-binding approximation, taking into account the
scattering by optical and acoustic phonons and by lateral disorder. It is
found that the distribution versus the in-plane kinetic energy depends
essentially on the ratio between the Bloch energy, $\varepsilon _{B}$, and
the optical phonon energy, $\hbar \omega _{0}$. The in-plane conductivity is
calculated for low-doped structures at temperatures 4.2 K and 20 K. The
negative conductivity is found for bias voltages corresponding to $%
\varepsilon _{B}/\hbar \omega _{0}\simeq $1/2, 1/3, 2/3$\ldots $ (the
Bloch-phonon resonance condition).
\end{abstract}

\maketitle

\section{Introduction}

Vertical charge transfer in a superlattice subjected to a homogeneous
electric field (biased superlattice, BSL, with equipopulated levels) has
been under investigation starting 70th (see Ref. \cite{1} and references in
the reviews of Ref. \cite{2}). The stimulated emission in the mid-infrared
(IR) and terahertz (THz) spectral regions, caused by the intersubband
transitions of electrons under vertical transport through tunnel-coupled
cascade structures (monopolar laser effect), has also been investigated.
Using this scheme, both mid-IR and THz lasers viability has been
demonstrated during the previous decade (see Refs.\cite{3,4} and references
therein ). Recently, nonequilibrium electron distribution has been observed
experimentally \cite{5} and described theoretically \cite{6} for
heavily-doped cascade structures, when the effective temperature is
determined from the balance equation. To the best of our knowledge, there is
no consideration of nonequilibrium carriers for low-doped structures which
were performed beyond the balance approach.

\begin{figure}[tbp]
\begin{center}
\includegraphics{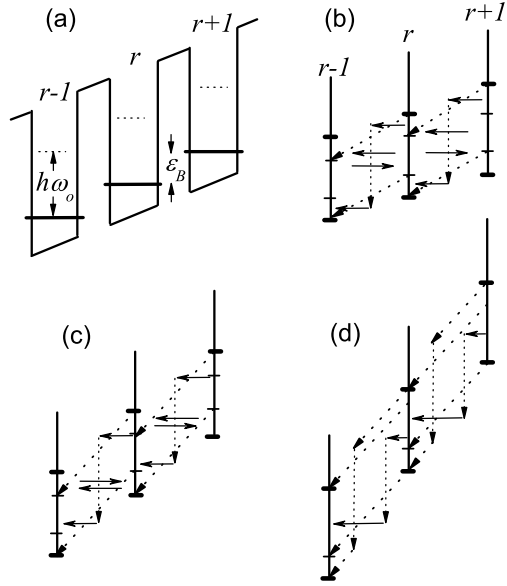}
\end{center}
\par
\addvspace{-2 cm}
\caption{(a) Band diagram for BSL of period $Z$ with the Bloch energy $%
\protect\varepsilon _{B}$ which is comparable to the optical phonon energy, $%
\hbar \protect\omega _{0}$. (b-d) Schemes of tunneling transitions due to
elastic scattering (solid arrows), spontaneous optical phonon emission
(dotted arrow), and phonon emission from the active region (vertical dashed
arrows) for the cases: (b) $\protect\varepsilon _{B}<\hbar \protect\omega %
_{0}/2$ , (c) $\hbar \protect\omega _{0}/2<\protect\varepsilon _{B}<\hbar 
\protect\omega _{0}$ , and (d) $\hbar \protect\omega _{0}<\protect%
\varepsilon _{B}$. }
\end{figure}

In this paper we study the nonequilibrium electron distribution in a biased
superlattice (BSL) under vertical current through the Wannier-Stark ladder,
which takes place under the condition $2T\ll \varepsilon _{B}$, where $%
\varepsilon _{B}$ is the Bloch energy and $T$ stands for the tunneling
matrix element between adjacent quantum wells (QWs) \cite{7}. Since the
parameters of each QW and the conditions for interwell transitions are
identical (see Fig. 1a) the level populations over QWs are the same. But the
distribution of electrons over the in-plane energy should change essentially
due to the interplay between elastic and non-elastic processes (see Figs.
1b-d). In the low-temperature case we only consider the passive region, with
energy less than the optical phonon energy, $\hbar \omega _{0}$. For the
low-concentration limit, we consider the kinetic equation which takes into
account the quasi-elastic scattering caused by acoustic phonons as well as
the interwell tunneling due to elastic scattering by disorder and due to
optical phonon emission. As a result, we obtain the electron distribution
versus the in-plane kinetic energy which strongly depends on the ratio $%
\varepsilon _{B}/\hbar \omega _{0}$. In the case of the Bloch-phonon
resonance, when $M\varepsilon _{B}=N\hbar \omega _{0}$ with $N$ and $M$
integers, \textit{a partially-inverted distribution}, with maxima at
energies $(N/M)\hbar \omega _{0}$, can be realized.

The phenomenon of absolute negative conductivity (ANC) of electrons excited
near the energy $\hbar \omega _{0}$ was discussed four decades ago \cite{8}
and different regimes of ANC (including magnetotransport\cite{9} and
transient regimes of the response\cite{10}) were considered. Recently, ANC
regime was observed when microwave radiation acts on two-dimensional (2D)
electrons in a quantizing magnetic field \cite{11}. As shown below, \textit{%
a resonant ANC regime} of the in-plane response can be obtained in BSL under
the Bloch-phonon resonance conditions. Such a peculiarity appears due to the
contribution of the energy interval near $\hbar \omega _{0}$ where an
inverted distribution takes place. As a result, BSL becomes instable with
respect to in-plane fluctuations if ANC conditions are satisfied.

The paper is organized as follows. The basic equations describing
distribution of hot electrons in BSL and in-plane conductivity are presented
in Sec. II. Analytical consideration for the case $\varepsilon _{B}/\hbar
\omega _{0}=1/2$ (the second-order Bloch-phonon resonance) is performed in
Sec. III. Results of numerical calculations are discussed in Sec. IV.
Concluding remarks and the list of assumptions made are given in the last
section. In Appendix, the kinetic equations for different $\varepsilon
_{B}/\hbar \omega _{0}$ are presented.

\section{Basic equations}

Within the tight-binding approximation, the electrons in BSL are
characterized by the 2D momentum, $\mathbf{p}$, and the quantum well number, 
$r=0,\pm 1,\ldots $. Under the in-plane electric field $\mathbf{E}$, the
distribution function, $f_{r\mathbf{p}}$, is governed by the system of
kinetic equations: 
\begin{equation}
e\mathbf{E}\cdot \frac{\partial f_{r\mathbf{p}}}{\partial \mathbf{p}}%
=\sum\limits_{k}J_{k}(f|r\mathbf{p}),  \label{1}
\end{equation}%
where the collision integrals $J_{k}(f|r\mathbf{p})$ describe the scattering
processes caused by the longitudinal optical phonon emission ($k=LO$), the
acoustic phonons ($k=ac$), or the static disorder ($k=d$). Below we present
these collision integrals and consider the kinetic equation for the
distribution functions $f_{r\mathbf{p}}$, which are normalized by the
condition $n=(2/V)\sum_{r\mathbf{p}}f_{r\mathbf{p}}$, where $n$ is the 3D
concentration, $V$ is the normalization volume, and the factor 2 is due to
spin. We also evaluate the lateral current density $\mathbf{I}=(2e/Vm)\sum_{r%
\mathbf{p}}\mathbf{p}f_{r\mathbf{p}}$ for electrons, with the effective mass 
$m$, under a weak probe field $\mathbf{E}$.

\subsection{Collision integrals}

Here we evaluate the collision integrals in Eq. (1) by modifying the general
expressions \cite{12} for electrons in BSL, described in the tight-binding
approximation by the states $|r\mathbf{p})$, with the energies $\varepsilon
_{rp}=r\varepsilon _{B}+\varepsilon _{p}$, where $\varepsilon _{p}=p^{2}/2m$
is the in-plane kinetic energy (see Sec. 3 in Ref. 7). For the
low-temperature case, when temperature of phonons $T_{ph}\ll \hbar \omega
_{0}$, the spontaneous emission of dispersionless optical phonons is
described by 
\begin{equation}
J_{LO}(f|r\mathbf{p})=\frac{2\pi }{\hbar }\sum\limits_{r^{\prime }\mathbf{p}%
^{\prime }\mathbf{Q}}|C_{Q}^{(LO)}|^{2}|(r^{\prime }\mathbf{p}^{\prime }|e^{i%
\mathbf{Qr}}|r\mathbf{p})|^{2}~\left[ \delta (\varepsilon _{r^{\prime
}p^{\prime }}-\varepsilon _{rp}-\hbar \omega _{0})f_{r^{\prime }\mathbf{p}%
^{\prime }}-\delta (\varepsilon _{rp}-\varepsilon _{r^{\prime }p^{\prime
}}-\hbar \omega _{0})f_{r\mathbf{p}}\right] ,  \label{2}
\end{equation}%
where $|C_{Q}^{(LO)}|^{2}$ is the bulk matrix element for the Fr\"{o}lich
interaction, with the vibration mode characterized by the 3D wave vector $%
\mathbf{Q}$, and $|(r^{\prime }\mathbf{p}^{\prime }|e^{i\mathbf{Qr}}|r%
\mathbf{p})|^{2}$ is the overlap factor. Taking into account the
quasielastic energy relaxation caused by the equipopulated acoustic phonons,
one obtains the collision integral 
\begin{equation}
J_{ac}(f|r\mathbf{p})=\sum\limits_{r^{\prime }\mathbf{p}^{\prime }}W_{r%
\mathbf{p}r^{\prime }\mathbf{p}^{\prime }}(f_{r^{\prime }\mathbf{p}^{\prime
}}-f_{r\mathbf{p}})-\frac{1}{2}\sum\limits_{r^{\prime }\mathbf{p}^{\prime
}}\Delta W_{r\mathbf{p}r^{\prime }\mathbf{p}^{\prime }}(f_{r^{\prime }%
\mathbf{p}^{\prime }}+f_{r\mathbf{p}}).  \label{3}
\end{equation}%
The transition probabilities $W_{r\mathbf{p}r^{\prime }\mathbf{p}^{\prime }}$
and $\Delta W_{r\mathbf{p}r^{\prime }\mathbf{p}^{\prime }}$ are written here
within the second order accuracy with respect to the acoustic phonon energy, 
$\hbar \omega _{Q}$, as follows 
\begin{eqnarray}
W_{r\mathbf{p}r^{\prime }\mathbf{p}^{\prime }} &=&K_{rr^{\prime }}(\mathbf{p}%
-\mathbf{p}^{\prime })\delta (\varepsilon _{r^{\prime }p^{\prime
}}-\varepsilon _{rp})+\frac{T_{ph}}{2}\Delta K_{rr^{\prime }}(\mathbf{p}-%
\mathbf{p}^{\prime })\delta ^{\prime \prime }(\varepsilon _{r^{\prime
}p^{\prime }}-\varepsilon _{rp}),  \nonumber \\
\Delta W_{r\mathbf{p}r^{\prime }\mathbf{p}^{\prime }} &=&\Delta
K_{rr^{\prime }}(\mathbf{p}-\mathbf{p}^{\prime })\delta (\varepsilon
_{r^{\prime }p^{\prime }}-\varepsilon _{rp}).  \label{4}
\end{eqnarray}%
Here $\delta ^{\prime }(E)$ and $\delta ^{\prime \prime }(E)$ are the first
and second derivatives of the $\delta $-function and the kernels $%
K_{rr^{\prime }}$ and $\Delta K_{rr^{\prime }}$ are given by 
\begin{eqnarray}
K_{rr^{\prime }}(\mathbf{p}-\mathbf{p}^{\prime }) &=&\frac{4\pi }{\hbar }%
\sum\limits_{\mathbf{Q}}|C_{Q}^{(ac)}|^{2}|(r^{\prime }\mathbf{p}^{\prime
}|e^{i\mathbf{Qr}}|r\mathbf{p})|^{2}\frac{T_{ph}}{\hbar \omega _{Q}}~ 
\nonumber \\
\Delta K_{rr^{\prime }}(\mathbf{p}-\mathbf{p}^{\prime }) &=&\frac{4\pi }{%
\hbar }\sum\limits_{\mathbf{Q}}|C_{Q}^{(ac)}|^{2}|(r^{\prime }\mathbf{p}%
^{\prime }|e^{i\mathbf{Qr}}|r\mathbf{p})|^{2}\hbar \omega _{Q}  \label{5}
\end{eqnarray}%
where $C_{Q}^{(ac)}$ is the bulk matrix element for the deformation
interaction.

We restrict ourselves to the sequential tunneling processes under the
condition $T\ll \varepsilon _{B}$. Considering only the proportional to $%
(T/\varepsilon _{B})^{2}$ corrections to the overlap factors, we use 
\begin{equation}
|(r^{\prime }\mathbf{p}^{\prime }|e^{i\mathbf{Qr}}|r\mathbf{p})|^{2}\simeq
\delta _{\mathbf{p}^{\prime }\mathbf{p}+\hbar \mathbf{q}}\Psi _{q_{\bot }d}%
\left[ \delta _{rr^{\prime }}+\left( \frac{T}{\varepsilon _{B}}\right)
^{2}\left( \delta _{rr^{\prime }+1}+\delta _{rr^{\prime }-1}\right) \right] ,
\label{6}
\end{equation}%
where $\Psi _{q_{\bot }d}=|(0|e^{iq_{\bot }z}|0)|^{2}$ describes the
transverse overlap between the ground states of the QWs, $|0)$. Since all
QWs are identical and $\varepsilon _{r^{\prime }p^{\prime }}-\varepsilon
_{rp}=(r-r^{\prime })\varepsilon _{B}+\varepsilon _{p^{\prime }}-\varepsilon
_{p}$, the distribution functions are the same in any QW, i.e. $f_{r\mathbf{p%
}}\rightarrow f_{\mathbf{p}}$. Thus, the collision integrals in Eqs. (2) and
(3) are independent on $r$ because the summation over $r^{\prime }$ is
replaced by $\sum_{\Delta r=\pm 1}\ldots $. The collision integral in Eq.
(2) is transformed into 
\begin{eqnarray}
J_{LO}(f|\mathbf{p}) &\simeq &\frac{2\pi }{\hbar }\sum\limits_{r^{\prime }%
\mathbf{p^{\prime }}q_{\bot }}|C_{Q}^{(LO)}|^{2}\Psi _{q_{\bot }d}~~~~~ 
\nonumber \\
&&\times\biggr\{\delta (\varepsilon _{p^{\prime }}-\varepsilon _{p}-\hbar
\omega_{0})f_{\mathbf{p}^{\prime }}-\delta (\varepsilon _{p}-\varepsilon
_{p^{\prime }}-\hbar\omega _{0})f_{\mathbf{p}}  \nonumber \\
&&+\biggr(\frac{T}{\varepsilon _{B}}\biggr)^{2}\sum\limits_{\Delta r=\pm 1}%
\biggr[\delta (\Delta r\varepsilon _{B}+\varepsilon _{p^{\prime
}}-\varepsilon _{p}-\hbar \omega _{0})f_{\mathbf{p}^{\prime }}  \nonumber \\
&&-\delta (\Delta r\varepsilon _{B}+\varepsilon _{p^{\prime }}-\varepsilon
_{p}+\hbar \omega _{0})f_{\mathbf{p}}\biggr]\biggr\},  \label{7}
\end{eqnarray}%
where $\sum_{\Delta r=\pm 1}\ldots $ describes the interwell tunneling with
LO-phonon emission and $Q^{2}=|\mathbf{p}-\mathbf{p}^{\prime }|^{2}/\hbar
^{2}+q_{\bot }^{2}$.

Below we restrict ourselves to the thin QW case, when $|C_{Q}^{(ac)}|^{2}$
can be replaced by $|C_{q_{\bot }}^{(ac)}|^{2}$. Similar transformations for
the acoustic phonon contribution of Eq. (3) give us 
\begin{eqnarray}
J_{ac}(f|\mathbf{p}) &=&K_{ac}\sum\limits_{\mathbf{p}^{\prime }}\biggr[%
\delta (\varepsilon _{p^{\prime }}-\varepsilon _{p})~+\sum\limits_{\Delta
r=\pm 1}\left( \frac{T}{\varepsilon _{B}}\right) ^{2}\delta (\Delta
r\varepsilon _{B}+\varepsilon _{p^{\prime }}-\varepsilon _{p})\biggr](f_{%
\mathbf{p}^{\prime }}-f_{\mathbf{p}})  \nonumber \\
&&-\Delta K\sum\limits_{\mathbf{p}^{\prime }}\frac{T_{ph}}{2}\delta ^{\prime
\prime }(\varepsilon _{p^{\prime }}-\varepsilon _{p})(f_{\mathbf{p}^{\prime
}}-f_{\mathbf{p}})-\frac{1}{2}\delta ^{\prime }(\varepsilon _{p^{\prime
}}-\varepsilon _{p})(f_{\mathbf{p}^{\prime }}+f_{\mathbf{p}}).~~~~~
\label{8}
\end{eqnarray}%
Here we have neglected weak ($\propto \Delta K$) contributions to the
tunneling transitions. The kernels in Eq. (5) appear to be momentum
independent 
\begin{eqnarray}
K_{ac} &\approx &\frac{4\pi }{\hbar }\sum\limits_{q_{\bot }}|C_{q_{\bot
}}^{(ac)}|^{2}\Psi _{q_{\bot }d}\frac{T_{ph}}{\hbar \omega _{q_{\bot }}} 
\nonumber \\
\Delta K &\approx &\frac{4\pi }{\hbar }\sum\limits_{q_{\bot }}|C_{q_{\bot
}}^{(ac)}|^{2}\Psi _{q_{\bot }d}\hbar \omega _{q_{\bot }}  \label{9}
\end{eqnarray}%
due to the narrow QW approximation. The intra- and inter-well scattering
caused by the static disorder can be described in a similar way to the
elastic ($\propto K_{ac}$) contributions in Eq. (8): 
\begin{equation}
J_{d}(f|\mathbf{p})=K_{d}\sum\limits_{\mathbf{p}^{\prime }}\biggr[\delta
(\varepsilon _{p^{\prime }}-\varepsilon _{p})~+\sum\limits_{\mathbf{p}%
^{\prime }\Delta r=\pm 1}\left( \frac{T}{\varepsilon _{B}}\right) ^{2}\delta
(\Delta r\varepsilon _{B}+\varepsilon _{p^{\prime }}-\varepsilon _{p})%
\biggr] (f_{\mathbf{p}^{\prime }}-f_{\mathbf{p}})  \label{10}
\end{equation}%
Factors $K_{d}$ and $K_{ac}$ determine the departure relaxation rates caused
by the elastic scattering mechanisms as $\nu _{d,ac}=K_{d,ac}\sum_{\mathbf{p}%
^{\prime }}\delta (\varepsilon _{p^{\prime }}-\varepsilon _{p})\propto \rho
_{2D}$, where $\rho _{2D}$ is the 2D density of states.

\subsection{Nonequilibrium distribution}

We search for the solution of Eq. (1) in the form $f_{\mathbf{p}}\simeq
f_{\varepsilon }+\Delta f_{\mathbf{p}}$, where $f_{\varepsilon }$ describes
the lateral heating due to tunneling current and $\Delta f_{\mathbf{p}}$ is
the in-plane anisotropic addendum due to the weak field $\mathbf{E}$. We
consider the symmetric part of the distribution which is governed by the
kinetic equation $\sum_{k}J_{k}(f|\varepsilon )=0$ and satisfies the
normalization condition 
\begin{equation}
nZ=\rho _{2D}\int_{0}^{\infty }d\varepsilon f_{\varepsilon }  \label{11}
\end{equation}%
with the layer concentration $nZ$. Averaging Eq. (7) over $\mathbf{p}$-plane
and taking into account the energy conservation condition, one obtains the
LO-contribution as a finite-difference form: 
\begin{equation}
J_{LO}(f|\varepsilon )=\nu _{\varepsilon +\hbar \omega _{0},\varepsilon
}f_{\varepsilon +\hbar \omega _{0}}-\nu _{\varepsilon -\hbar \omega
_{0},\varepsilon }f_{\varepsilon }~+\sum\limits_{\Delta r=\pm 1}[\nu
_{\varepsilon +\hbar \omega _{0}-\Delta r\varepsilon _{B},\varepsilon
}^{t}f_{\varepsilon +\hbar \omega _{0}-\Delta
r\varepsilon_{B}}-\nu_{\varepsilon -\hbar \omega _{0}-\Delta r\varepsilon
_{B},\varepsilon }^{t}f_{\varepsilon }].  \label{12}
\end{equation}%
Here the tunneling contributions $\nu _{E,\varepsilon }^{t}=(T/\varepsilon
_{B})^{2}\nu _{E,\varepsilon }$ are reduced by the factor $(T/\varepsilon
_{B})^{2}$. We introduce the relaxation rate describing the spontaneous
emission of LO-phonons as follows 
\begin{equation}
\nu _{E,\varepsilon }=\frac{2\pi }{\hbar }\sum\limits_{\mathbf{p}^{\prime
}q_{\bot }}|C_{Q}^{(LO)}|^{2}\Psi _{q_{\bot }d}\delta (\varepsilon
_{p^{\prime }}-E).  \label{13}
\end{equation}%
Performing the integration over $\mathbf{p}^{\prime }$-plane one obtains 
\begin{equation}
\nu _{E,\varepsilon }=\theta (E)\alpha \omega _{o}\int\limits_{-\infty
}^{\infty }dx\Psi _{x}\sqrt{\frac{\varepsilon _{d}\hbar \omega _{o}}{%
(\varepsilon _{d}x^{2}+\varepsilon +E)^{2}-4\varepsilon E}},  \label{14}
\end{equation}%
where $\alpha $ is the polaron coupling constant and $\varepsilon
_{d}=(\hbar /d)^{2}/2m$. This rate is of the order of $\alpha \omega _{o}$.
Fig. 2 shows the dimensionless relaxation rate $\nu _{E,\varepsilon }/\alpha
\omega _{0}$ versus $\varepsilon /\hbar \omega _{o}$, plotted in the passive
region for different $E/\hbar \omega _{o}$\ values and \ for a 60 \AA\ wide
QW, when $\varepsilon _{d}/\hbar \omega _{o}\simeq $0.44. Note, that $\nu
_{E,\varepsilon }$ appears to be logarithmically divergent if $\
E\rightarrow \varepsilon $.

\begin{figure}[tbp]
\begin{center}
\includegraphics{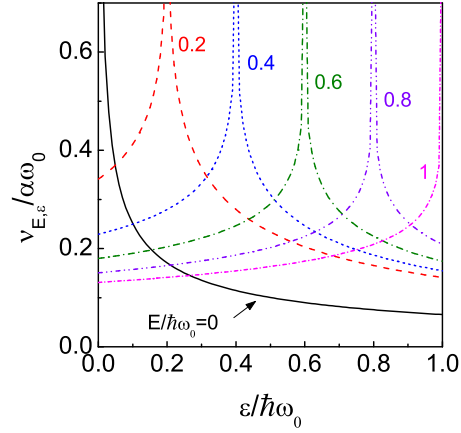}
\end{center}
\par
\addvspace{-4 cm}
\caption{(Color online) Dimensionless relaxation rate $\protect\nu _{E,%
\protect\varepsilon }/\protect\alpha \protect\omega _{0}$ versus $\protect%
\varepsilon /\hbar \protect\omega _{0}$ plotted in the passive region for $%
E/\hbar \protect\omega _{0}=$ 0, 0.2, 0.4, 0.6, 0.8 and 1. }
\end{figure}

The intrawell process of quasi-elastic energy relaxation is described by the
Fokker-Planck collision integral \cite{12} given by 
\begin{equation}
J_{ac}(f|\varepsilon )\approx \nu _{ac}\bar{\varepsilon}^{2}\frac{d}{%
d\varepsilon }\left( \frac{df_{\varepsilon }}{d\varepsilon }+\frac{%
f_{\varepsilon }}{T_{ph}}\right) ,  \label{15}
\end{equation}%
where $\nu _{ac}$ is the above-introduced departure relaxation rate and $%
\bar{\varepsilon}^{2}\simeq (\Delta K/K_{ac})T_{ph}/2$. The elastic
tunneling relaxation caused by disorder and acoustic phonons [see \ Eqs. (8)
and (10)] is governed by the finite-difference contribution 
\begin{equation}
J_{t}(f|\varepsilon )\simeq \nu _{t}\sum\limits_{\Delta r=\pm 1}\theta
(\varepsilon -\Delta r\varepsilon _{B})(f_{\varepsilon -\Delta r\varepsilon
_{B}}-f_{\varepsilon })  \label{16}
\end{equation}%
with the tunneling rate $\nu _{t}=(T/\varepsilon _{B})^{2}(\nu _{d}+\nu
_{ac})$. Thus, the distribution $f_{\varepsilon }$ is governed by the
equation 
\begin{equation}
J_{ac}(f|\varepsilon )+J_{t}(f|\varepsilon )+J_{LO}(f|\varepsilon )=0.
\label{17}
\end{equation}%
Moreover, in the active region, $\varepsilon >\hbar \omega _{o}$, the main
contribution is due to the spontaneous emission of LO-phonons [first and
second terms of Eq. (12)].

In the active region, $\varepsilon >\hbar \omega _{0}$, the distribution
decreases fast. Thus, in the kinetic equation, one has to take into account
the second derivative from Eq. (15) and the spontaneous emission
contribution form Eq. (12): 
\begin{equation}
\nu _{ac}\bar{\varepsilon}^{2}\frac{d^{2}f_{\varepsilon }}{d\varepsilon ^{2}}%
-\nu _{LO}f_{\varepsilon }=0,  \label{18}
\end{equation}%
where $\nu _{LO}=\nu _{0,\hbar\omega _{0}}$. Using the boundary condition $%
f_{\varepsilon \rightarrow \infty }=0$ one obtains the solution $%
f_{\varepsilon }\approx f_{\hbar \omega _{0}}\exp [-\lambda _{0}(\varepsilon
-\hbar \omega _{0})]$ with $\lambda _{0}=\sqrt{\nu _{LO}/\nu _{ac}}\bar{%
\varepsilon}^{-1}$.

Next, eliminating the fast spontaneous emission of LO-phonons one obtains
the Eq. (17) in the passive region with the boundary condition 
\begin{equation}
\left( {\frac{d}{{d\varepsilon }}+\lambda _{0}}\right) f_{\varepsilon
\rightarrow \hbar \omega _{0}-0}=0.  \label{19}
\end{equation}%
Thus, the problem formulated in the passive region takes into account the
quasi-elastic energy relaxation described by Eq. (15) and the interwell
tunneling transitions shown in Figs. 1b-d. The normalization condition
should be restricted over the passive region and Eq. (11) is transformed
into $nZ=\rho _{2D}\int_{0}^{\hbar \omega _{0}}d\varepsilon f_{\varepsilon }$%
.

\subsection{Lateral conductivity}

Further, we turn to the description of the linear response given by $\Delta
f_{\mathbf{p}}\propto \mathbf{E}$ and consider the current density 
\begin{equation}
\mathbf{I}=\frac{2e}{Zm}\int \frac{d\mathbf{p}}{(2\pi \hbar )^{2}}\mathbf{p}%
\Delta f_{\mathbf{p}}.  \label{20}
\end{equation}%
The nonsymmetric part of the distribution function $\Delta f_{\mathbf{p}}$
is determined from the linearized kinetic equation 
\begin{equation}
e\mathbf{E}\cdot \frac{\partial f_{\varepsilon }}{\partial \mathbf{p}}%
=\sum\limits_{k}J_{k}(\Delta f|\mathbf{p})  \label{21}
\end{equation}%
where the elastic collision integrals due to $ac$- and $d$-contributions can
be replaced by $-\nu _{m}\Delta f_{\mathbf{p}}$. Here $\nu _{m}=\nu _{d}+\nu
_{ac}$ is the momentum relaxation rate due to the elastic scattering [see
Eqs. (8) and (10)]. The non-elastic momentum relaxation due to the
optical-phonon-induced interwell transitions, $J_{LO}(\Delta f|\mathbf{p})$,
is given by the $\propto (T/\varepsilon _{B})^{2}$ contribution of Eq. (7).
Introducing the energy-dependent function $\chi _{\varepsilon }$ according
to $\Delta f_{\mathbf{p}}=(e/m)(\mathbf{E}\cdot \mathbf{p})\chi
_{\varepsilon }$, we transform Eq. (21) into the finite-difference equation 
\begin{equation}
\frac{df_{\varepsilon }}{d\varepsilon }=-\nu _{m}\chi _{\varepsilon
}+\sum\limits_{\Delta r=\pm 1}[\widetilde{\nu }_{\varepsilon +\hbar \omega
_{0}-\Delta r\varepsilon _{B},\varepsilon }^{t}\chi _{\varepsilon +\hbar
\omega _{0}-\Delta r\varepsilon _{B}}-\nu _{\varepsilon -\hbar \omega
_{0}-\Delta r\varepsilon _{B},\varepsilon }^{t}\chi _{\varepsilon }].
\label{22}
\end{equation}%
Here $\widetilde{\nu }_{E,\varepsilon }^{t}=(T/\varepsilon _{B})^{2}%
\widetilde{\nu }_{E,\varepsilon }$ is the tunneling-induced relaxation rate
where 
\begin{equation}
\widetilde{\nu }_{E,\varepsilon }=\frac{2\pi }{\hbar }\sum\limits_{\mathbf{p}%
^{\prime }q_{\bot }}|C_{Q}^{(LO)}|^{2}\Psi _{q_{\bot }d}\cos (\widehat{%
\mathbf{p},\mathbf{p}^{\prime }})\delta (\varepsilon -\varepsilon
_{p^{\prime }}-E)  \label{23}
\end{equation}%
which uses a similar notation to Eq. (13).

Introducing the in-plane conductivity, $\sigma $, according to $\mathbf{I}%
=\sigma \mathbf{E}$ we obtain 
\begin{equation}
\sigma =\frac{e^{2}\rho _{2D}}{mZ}\int\limits_{0}^{\hbar \omega
_{0}}d\varepsilon \varepsilon \chi _{\varepsilon }.  \label{24}
\end{equation}%
Here we neglect the contribution from $\varepsilon >\hbar \omega _{0}$
because of the smallness of $\chi _{\varepsilon }\simeq \nu
_{LO}^{-1}(-df_{\varepsilon }/d\varepsilon )$ in the active region. Under
the condition $\nu _{m}\gg \nu _{LO}^{t}$ Eq. (22) gives $\chi _{\varepsilon
}\simeq \nu _{m}^{-1}(-df_{\varepsilon }/d\varepsilon )$ and the
conductivity takes the form 
\begin{equation}
\sigma =\sigma _{0}\left( 1-\frac{\rho _{2D}\hbar \omega _{0}}{n_{2D}}%
f_{\hbar \omega _{0}}\right) ,~~~~~~\sigma _{0}=\frac{e^{2}n_{2D}}{m\nu _{m}Z%
}.  \label{25}
\end{equation}%
As a result, $\sigma /\sigma _{0}$ is expressed through the distribution
function at $\varepsilon =\hbar \omega _{0}$ and a negative lateral
conductivity takes place at $f_{\hbar \omega _{0}}>n_{2D}/\rho _{2D}\hbar
\omega _{0}$.

\section{Second order Bloch-phonon resonance}

Before analyzing the general problem, we consider the simple resonant case $%
2\varepsilon _{B}=\hbar \omega _{0}$, when the distribution can be
considered over two intervals: $0<\varepsilon <\hbar \omega _{0}/2$ and $%
\hbar \omega _{0}/2<\varepsilon <\hbar \omega _{0}$. Introducing the
functions over the interval $0<\varepsilon <\hbar \omega _{0}/2$ according
to $f_{1\varepsilon }=f_{\varepsilon }$ and $f_{2\varepsilon
}=f_{\varepsilon +\hbar \omega _{0}}$, we transform Eq. (17) [or Eqs. (A1)
and (A3) in Appendix] into the system 
\begin{eqnarray}
J_{ac}(f_{1}|\varepsilon )+\nu _{t}\left( 2f_{2\varepsilon }-f_{1\varepsilon
}\right) +\nu _{\varepsilon +\hbar \omega _{0}/2,\varepsilon
}^{t}f_{2\varepsilon } = 0  \label{26} \\
J_{ac}(f_{2}|\varepsilon )+\nu _{t}\left( f_{1\varepsilon }-2f_{2\varepsilon
}\right) -\nu _{\varepsilon ,\varepsilon +\hbar \omega
_{0}/2}^{t}f_{2\varepsilon } = 0  \nonumber
\end{eqnarray}%
Here $\nu _{t}\simeq (T/\varepsilon _{B})^{2}\nu _{m}$ is the elastic
tunneling rate and $\nu _{E,\varepsilon }^{t}$ was introduced in Eq. (12).
The two second-order differential equations (26) should be solved with the
boundary condition of Eq. (19), the normalization condition $nZ=\rho
_{2D}\int_{0}^{\hbar \omega _{0}/2}d\varepsilon (f_{1\varepsilon
}+f_{2\varepsilon })$, as well as the inhomogeneous conditions $f_{1\hbar
\omega _{0}/2}=f_{2\varepsilon =0}$ and $(df_{1\varepsilon }/d\varepsilon
)_{\hbar \omega _{0}/2}=(df_{2\varepsilon }/d\varepsilon )_{0}$.

\begin{figure}[tbp]
\begin{center}
\includegraphics{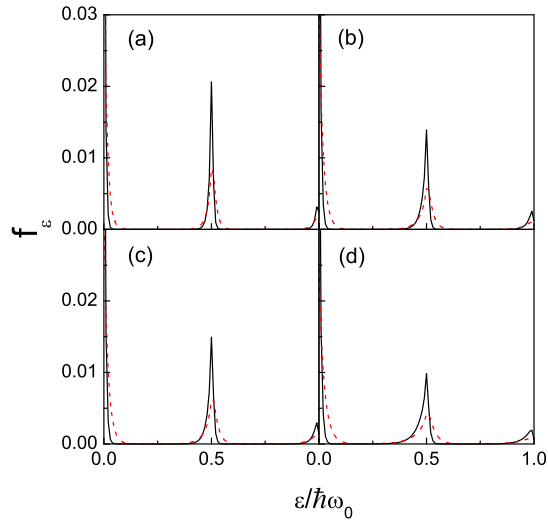}
\end{center}
\par
\addvspace{-4 cm}
\caption{(Color online) Distribution function $f_{\protect\varepsilon }$ vs
in-plane kinetic energy obtained from the system (26) at $T_{ph}=$4.2 K and
20 K (solid and dashed curves) for (a) $T$=5 meV and $\protect\nu _{m}$=1 ps$%
^{-1}$, (b) $T$=5 meV and $\protect\nu _{m}$=0.5 ps$^{-1}$, (c) $T$=3.5 meV
and $\protect\nu _{m}$=1 ps$^{-1}$, and (d) $T$=3.5 meV and $\protect\nu %
_{m} $=0.5 ps$^{-1}$. }
\end{figure}
Fig. 3 shows the distribution function $f_{\varepsilon }$ versus in-plane
kinetic energy obtained from the system (26) for different $T_{ph}$, $T$,
and $\nu _{m}$. As can be seen, peaks of $f_{\varepsilon }$ widen as
temperature increases, reducing their maxima. The effect of interwell
coupling $\propto T$ and elastic scattering $\propto \nu _{m}$ is also
evident, founding that peaks increase as these parameters do. Calculations
have been made for a concentration $n_{2D}=10^{9}$ cm$^{-2}$ (or $n=10^{15}$
cm$^{-3}$ if $Z=$100 \AA ), so that $f_{1\varepsilon }\leq 0.1$ and
electrons are non-degenerate. \cite{13} For this value, $n_{2D}/\rho
_{2D}\hbar \omega _{0}\sim 10^{-3}$. Peaks located at $\hbar \omega _{0}/2$
and $\hbar \omega _{0}$ are of the order of 10$^{-2}$and 10$^{-3}$,
respectively. Therefore, the last value is big enough to obtain ANC because
of $f_{\hbar \omega _{0}}>n_{2D}/\rho _{2D}\hbar \omega _{0}$, according to
Eq. (25). In order to magnify peaks at $\varepsilon /\hbar \omega _{0}$=0.5
and 1, we have limited vertical axis size.

\begin{figure}[tbp]
\begin{center}
\includegraphics{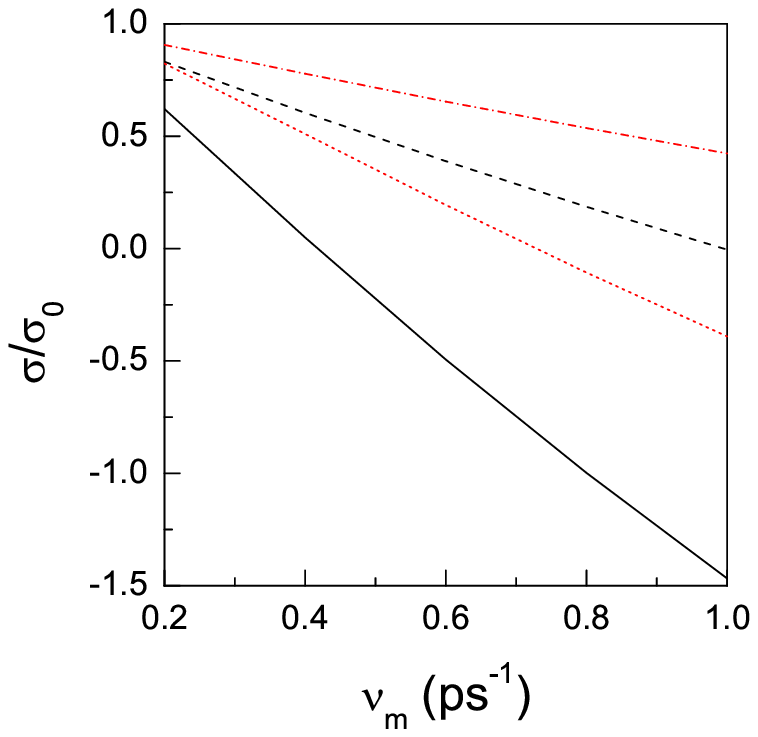}
\end{center}
\par
\addvspace{-4 cm}
\caption{(Color online) Normalized conductivity $\protect\sigma /\protect%
\sigma _{o}$, given by Eq. (25), vs momentum relaxation rate $\protect\nu %
_{m}$, for different temperatures and tunneling coupling values ($T_{ph}$
and $T$). Solid line: (4.2 K and 5 meV). Dashed line: (20 K and 5 meV).
Dotted line: (4.2 K and 3.5 meV). Dash-dotted line: (20 K and 3.5 meV). }
\end{figure}
Fig. 4 describes normalized lateral conductivity $\sigma /\sigma _{0}$ given
by Eq. (25) vs momentum relaxation rate $\nu _{m}$, for temperatures $%
T_{ph}= $4.2 K and 20 K, and tunneling coupling $T=$ 5 and 3.5 meV. Because
of the distribution function peaks at\ $\hbar \omega _{0}$ increase as $T$
and $\nu _{m}$ do (see Fig. 3), lateral conductivity decreases
correspondingly leading to negative values for a wide region of parameters.
As we saw in Fig. 3, the effect of the temperature is opposed to the
previous ones. For high temperatures $\sigma $ increases to reach $\sigma
_{0}$ and the possibility of having ANC disappears.

\section{Numerical results}

We turn now to the numerical calculations of the nonequilibrium distribution 
$f_{\varepsilon }$ governed by the Eq. (17), with the boundary condition
defined by Eq. (19), and the normalization requirement [the explicit form of
Eq. (17) for the cases $\varepsilon _{B}<\hbar \omega _{0}/2$, $\hbar \omega
_{0}/2<\varepsilon _{B}<\hbar \omega _{0}$, and $\hbar \omega
_{0}<\varepsilon _{B}$ are given in Appendix]. We also analyze the lateral
conductivity solving the finite-difference Eq. (22) [see explicit
expressions (A2), (A4), and (A6)] and performing the integration in Eq.
(24). Calculations below are performed for the GaAs/Al$_{0.3}$Ga$_{0.7}$%
As-based SL, formed by 60 \AA\ wide QWs separated by barriers of 32 \AA ~
(or 37 \AA )~ wide, which correspond to the tunneling matrix element $T$= 5
meV (or 3.5 meV). We consider temperatures of $T_{ph}=$4.2 K and 20 K as
well as the effect of the elastic scattering variation through the momentum
relaxation rate $\nu _{m}$=1 ps$^{-1}$ and 0.5 ps$^{-1}$. It is convenient
to use $n_{2D}=10^{9}$ cm$^{-2}$ in spite of $\sigma /\sigma_0$ does not
depend on concentration.

\begin{figure}[tbp]
\begin{center}
\includegraphics{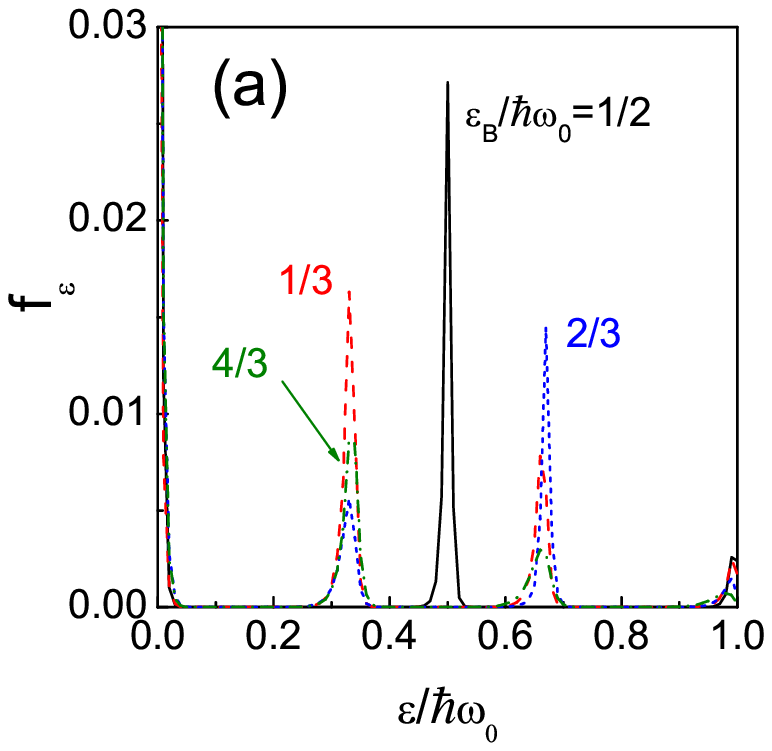} \includegraphics{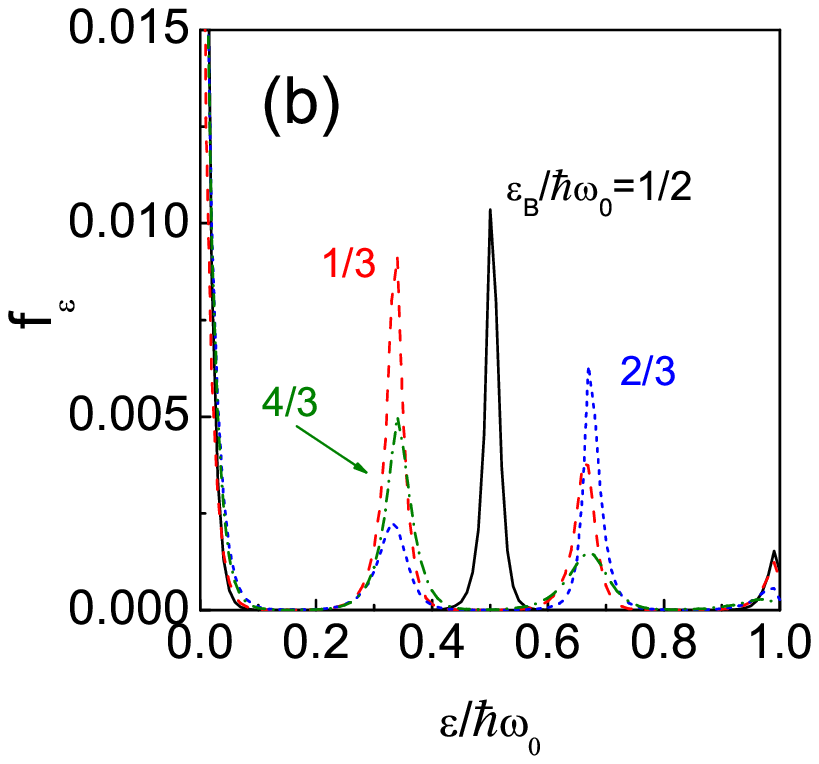}
\end{center}
\par
\addvspace{-4 cm}
\caption{ (Color online) Distribution function $f_{\protect\varepsilon }$ vs
in-plane kinetic energy for different $\protect\varepsilon _{B}/\hbar 
\protect\omega _{0}$ values (1/3, 1/2, 2/3, and 4/3) in GaAs/Al$_{0.3}$Ga$%
_{0.7}$As-based BSL at different temperatures (a) $T_{ph}$=4.2 K, and (b) $%
T_{ph}$=20 K. }
\end{figure}
Fig. 5 displays the distribution function $f_{\varepsilon }$ vs the in-plane
kinetic energy for different $\varepsilon _{B}/\hbar \omega _{o}$ values
(1/3, 1/2, 2/3, and 4/3) obtained from the general Eq. (17) (see also
Appendix). When comparing the case $\varepsilon _{B}/\hbar \omega _{o}=$1/2
of Fig. 5(a,b) with the panel (a) in Fig. 3, calculated in the former
section for the same parameters, a good agreement is found. As mentioned
before, temperature effect is reflected as a widening and decreasing of the
peaks. For other $N/M$ values $f_{\varepsilon }$ shows lower relative maxima.

\begin{figure}[tbp]
\begin{center}
\includegraphics{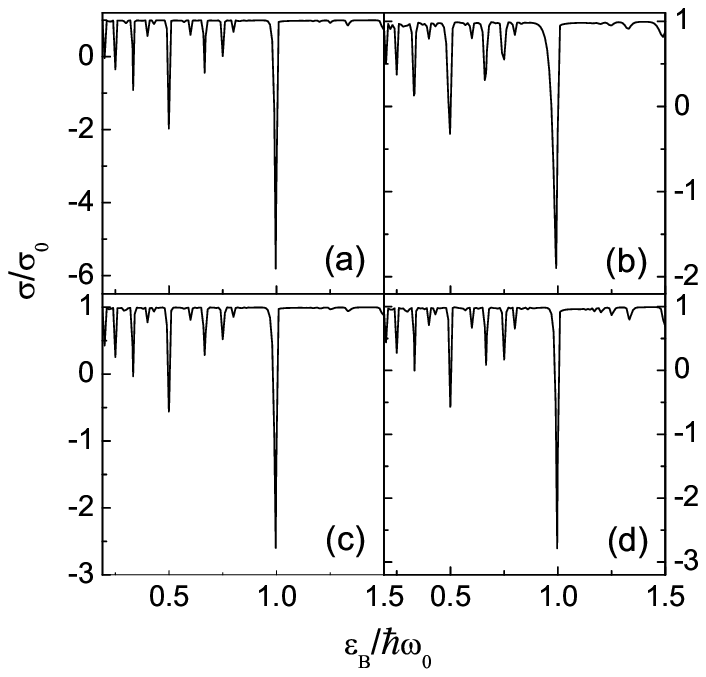}
\end{center}
\par
\addvspace{-4 cm}
\caption{Normalized lateral conductivity vs Bloch energy $\protect%
\varepsilon _{B}/\hbar \protect\omega _{0}$ and for different temperature,
elastic scattering and tunneling coupling ($T_{ph}$, $\protect\nu _{m}$, and 
$T$). $(a)$: (4.2 K, 1 ps$^{-1}$, and 5 meV). $(b)$: (20 K, 1 ps$^{-1}$, and
5 meV). $(c)$: (4.2 K, 1 ps$^{-1}$, and 3.5 meV). $(d)$: (4.2 K, 0.5 ps$%
^{-1} $, and 5 meV). }
\end{figure}
Next we calculate the normalized lateral conductivity solving Eq. (22) and
using $f_{\varepsilon }$ obtained before. Fig. 6 represents $\sigma /\sigma
_{0}$ as function of the Bloch energy $\varepsilon _{B}/\hbar \omega _{0}$
and for different temperature, elastic scattering and tunneling coupling
values. General behavior shows a pronounced relative minimum located at $%
\varepsilon _{B}/\hbar \omega _{0}=$1, followed by other relative minima at
1/2, 1/3, \ 2/3, 1/4,... (in decreasing order). In the active region, when $%
N>M$, these peaks are practically negligible. Depending on parameters some
of the peaks reach negative values. To clarify the effect of these
parameters we can compare in pairs the panels in Fig. 6. Thus, comparing
panels (a) and (b) we can see the temperature effect, which is similar to
the one found before for the distribution function: peaks are wider and less
pronounced, leading to less negative values. Comparing panels (a) and (c)
the effect of the tunneling coupling can be visible: if $T$ decreases
(increasing barriers width) size of peak minima are reduced. Finally, an
analogous behavior is observed comparing panels (a) and (d) to see the
elastic scattering effect. Reducing $\nu _{m}$, we obtain a decreasing of
the peaks in a similar way. One can conclude that the most favorable
conditions to get negative conductivity correspond to low temperature, and
big tunneling coupling and elastic scattering values.

\begin{figure}[tbp]
\begin{center}
\includegraphics{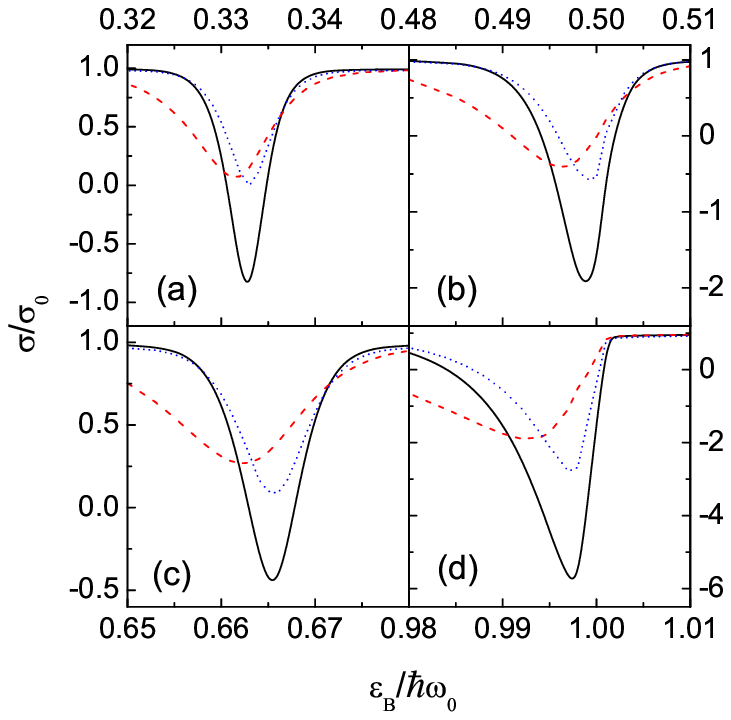}
\end{center}
\par
\addvspace{-4 cm}
\caption{(Color online) Normalized conductivity vs Bloch energy $\protect%
\varepsilon _{B}/\hbar \protect\omega _{0}$ around 1/3 (a), 1/2 (b), 2/3
(c), and 1 (d), for $T=$5 meV. Solid lines: $\protect\nu _{m}$=1 ps$^{-1}$
and $T_{ph}=$4.2 K. Dashed lines: $\protect\nu _{m}$=1 ps$^{-1}$ and $%
T_{ph}= $20 K. Dotted lines: $\protect\nu _{m}$=0.5 ps$^{-1}$ and $T_{ph}=$%
4.2 K.}
\end{figure}
In order to detail the shape of the normalized conductivity peaks we present
some of them in Fig. 7. We have chosen the more noticeable ones,
corresponding to Bloch energy $\varepsilon _{B}/\hbar \omega _{0}$ around
1/3 [panel (a)], 1/2 (b), 2/3 (c), and 1 (d), for $T=$5 meV, with
temperature and elastic scattering values corresponding to panels (a), (b)
and (d) included in Fig. 6. A breakdown of the peak symmetry is observable
when $\varepsilon _{B}/\hbar \omega _{0}$ increases going from a quite
symmetric peak for $\varepsilon _{B}/\hbar \omega _{0}$ close to 1/3, to a
clearly asymmetric peak around 1.

\section{Concluding remarks}

In summary, we have demonstrated that the negative lateral conductivity
regime is possible in low-doped biased superlattices under the Bloch-phonon
resonance conditions. When analyzing the dependence of $\sigma $ vs bias
voltage, narrow negative peaks take place if $\varepsilon _{B}/\hbar \omega
_{0}$ is close to the ratio 1/2, 1/3, 2/3$\ldots $. ANC regime appears to be
most pronounced for low-temperature region in BSL with an effective
interwell coupling.

Next, we list the assumptions used. The main restriction of the ANC regime
consists in neglecting the electron-electron scattering, imposed by the
Maxwell distribution, with an effective temperature suppressing the
high-energy part of the distribution. The condition $\sigma <0$ does not
depend on concentration for non-degenerate electrons. Thus, we have used in
the calculations $n_{2D}<10^{10}$ cm$^{-2}$, where the electron-electron
scattering is not the main scattering process (see Ref. 14, where different
systems were considered). The evaluation of the limiting concentration,
which requires to involve the electron-electron collision integral in Eq.
(17), lies beyond the scope of this paper. Another approximation we have
made is rather standard. In order to estimate the coefficients of the
kinetic equation, we have used a tight-binding approach for the description
of the electronic states\cite{7}. The use of the bulk model for phonon
dispersion and electron-phonon interaction is a reasonable approximation for
the GaAs/AlGaAs-based structures\cite{15}. We consider the momentum
relaxation due to short-range scattering neglecting a large-scale potential;
last contributions require a special attention in analogy with the case of a
single low-doped well. We restrict ourselves to the case of uniform bias
fields and equipopulated wells neglecting the possible domain formation (one
can avoid instabilities of vertical current in a short enough BSL\cite{16}).

We should also mention that an experimental task for measure the lateral
conductivity of BSL is not simple because a complicate contribution of the
corner regions is possible. But, instead of the dc current measurements, one
can use a high-frequency contactless study of the response using a
transverse capacitor. In addition, under the instability conditions (if $%
\sigma <0$) a direct measurement of lateral conductivity is not necessary
because the vertical current appears to be unstable. A detailed description
of this unstable response requires a special consideration.

To conclude, a low-doped BSL at low temperature is a suitable structure in
order to realize the absolute negative conductivity regime. In addition, a
similar behavior is possible not only for the BSL under consideration but
also for the more complicated tunnel-coupled structure used in quantum
cascade lasers. An instability of such a structure for the case of low
doping and temperature is possible and should be checked additionally.

\appendix

\section{Kinetic equations}

Below we rewrite Eqs. (17) and (22) for the cases $\varepsilon _{B}<\hbar
\omega _{0}/2$ (A), $\hbar \omega _{0}/2<\varepsilon _{B}<\hbar \omega _{0}$
(B), and $\hbar \omega _{0}<\varepsilon _{B}$ (C).

For the case A, when $\varepsilon _{B}<\hbar \omega _{0}-\varepsilon
_{B}<\hbar \omega _{0}$ (see Fig. 1b), the distribution in the passive
region $f_{\varepsilon }$ is described by the system 
\begin{equation}
\begin{array}{ll}
J_{ac}(f|\varepsilon )+\nu _{t}(f_{\varepsilon +\varepsilon
_{B}}+f_{\varepsilon +\hbar \omega _{0}-\varepsilon _{B}}-f_{\varepsilon
})+\nu _{\varepsilon +\hbar \omega _{0}-\varepsilon _{B},\varepsilon
}^{t}f_{\varepsilon +\hbar \omega _{0}-\varepsilon _{B}}=0,~ & 0<\varepsilon
<\varepsilon _{B} \\ 
J_{ac}(f|\varepsilon )+\nu _{t}(f_{\varepsilon +\varepsilon
_{B}}+f_{\varepsilon +-\varepsilon _{B}}-2f_{\varepsilon })=0,~ & 
\varepsilon _{B}<\varepsilon <\hbar \omega _{0}-\varepsilon _{B} \\ 
J_{ac}(f|\varepsilon )+\nu _{t}(f_{\varepsilon -\varepsilon
_{B}}-2f_{\varepsilon })-\nu _{\varepsilon -\hbar \omega _{0}+\varepsilon
_{B},\varepsilon }^{t}f_{\varepsilon }=0,~ & ~\hbar \omega _{0}-\varepsilon
_{B}<\varepsilon <\hbar \omega _{0}%
\end{array}
\label{A1}
\end{equation}%
which is written for the three intervals $0<\varepsilon <\varepsilon _{B}$, $%
\varepsilon _{B}<\varepsilon <\hbar \omega _{0}-\varepsilon _{B}$, and $%
\hbar \omega _{0}-\varepsilon _{B}<\varepsilon <\hbar \omega _{0}$. Eq.(22)
in this case is transformed into the system 
\begin{equation}
\begin{array}{ll}
\frac{df_{\varepsilon }}{d\varepsilon }=-\nu _{m}\chi _{\varepsilon }+%
\widetilde{\nu }_{\varepsilon +\hbar \omega _{0}-\varepsilon
_{B},\varepsilon }^{~t}\chi _{\varepsilon +\hbar \omega _{0}-\varepsilon
_{B}}, & 0<\varepsilon <\varepsilon _{B} \\ 
\frac{df_{\varepsilon }}{d\varepsilon }=-\nu _{m}\chi _{\varepsilon }, & 
\varepsilon _{B}<\varepsilon <\hbar \omega _{0}-\varepsilon _{B} \\ 
\frac{df_{\varepsilon }}{d\varepsilon }=-(\nu _{m}+\nu _{\varepsilon -\hbar
\omega _{0}+\varepsilon _{B},\varepsilon }^{t})\chi _{\varepsilon }, & \hbar
\omega _{0}-\varepsilon _{B}<\varepsilon <\hbar \omega _{0}%
\end{array}
\label{A2}
\end{equation}

In the case B, when $\hbar \omega _{0}-\varepsilon _{B}<\varepsilon
_{B}<\hbar \omega _{0}$ see Fig. 1c, Eq. (17) is transformed into the system 
\begin{equation}
\begin{array}{ll}
J_{ac}(f|\varepsilon )+\nu _{t}(f_{\varepsilon +\varepsilon
_{B}}+f_{\varepsilon +\hbar \omega _{0}-\varepsilon _{B}}-f_{\varepsilon
})+\nu _{\varepsilon +\hbar \omega _{0}-\varepsilon _{B},\varepsilon
}^{t}f_{\varepsilon +\hbar \omega _{0}-\varepsilon _{B}}=0, & 0<\varepsilon
<\hbar \omega _{0}-\varepsilon _{B} \\ 
J_{ac}(f|\varepsilon )+\nu _{t}(f_{\varepsilon +\hbar \omega
_{0}-\varepsilon _{B}}-f_{\varepsilon })+\nu _{\varepsilon +\hbar \omega
_{0}-\varepsilon _{B},\varepsilon }^{t}f_{\varepsilon +\hbar \omega
_{0}-\varepsilon _{B}}-\nu _{\varepsilon -\hbar \omega _{0}+\varepsilon
_{B},\varepsilon }^{t}f_{\varepsilon }=0, & \hbar \omega _{0}-\varepsilon
_{B}<\varepsilon <\varepsilon _{B} \\ 
J_{ac}(f|\varepsilon )+\nu _{t}(f_{\varepsilon -\varepsilon
_{B}}-2f_{\varepsilon })-\nu _{\varepsilon -\hbar \omega _{0}+\varepsilon
_{B},\varepsilon }^{t}f_{\varepsilon }=0, & \varepsilon _{B}<\varepsilon
<\hbar \omega _{0}%
\end{array}
\label{A3}
\end{equation}%
while Eq. (22) takes the form 
\begin{equation}
\begin{array}{ll}
\frac{df_{\varepsilon }}{d\varepsilon }=-\nu _{m}\chi _{\varepsilon }+%
\widetilde{\nu }_{\varepsilon +\hbar \omega _{0}-\varepsilon
_{B},\varepsilon }\chi _{\varepsilon +\hbar \omega _{0}-\varepsilon _{B}}, & 
0<\varepsilon <\hbar \omega _{0}-\varepsilon _{B} \\ 
\frac{df_{\varepsilon }}{d\varepsilon }=-(\nu _{m}+\nu _{\varepsilon -\hbar
\omega _{0}+\varepsilon _{B},\varepsilon }^{t})\chi _{\varepsilon }+%
\widetilde{\nu }_{\varepsilon +\hbar \omega _{0}-\varepsilon
_{B},\varepsilon }\chi _{\varepsilon +\hbar \omega _{0}-\varepsilon _{B}},~~
& ~\hbar \omega _{0}-\varepsilon _{B}<\varepsilon <\varepsilon _{B} \\ 
\frac{df_{\varepsilon }}{d\varepsilon }=-(\nu _{m}+\widetilde{\nu }%
_{\varepsilon -\hbar \omega _{0}+\varepsilon _{B},\varepsilon })\chi
_{\varepsilon }, & ~\varepsilon _{B}<\varepsilon <\hbar \omega _{0}%
\end{array}
\label{A4}
\end{equation}

Similarly, for high-biased SL (case C, when $\hbar \omega _{0}<\varepsilon
_{B}$ see Fig. 1d) one obtains the system 
\begin{equation}
\begin{array}{ll}
J_{ac}(f|\varepsilon )+\nu _{\varepsilon +2\hbar \omega _{0}-\varepsilon
_{B},\varepsilon }^{t}f_{\varepsilon +2\hbar \omega _{0}-\varepsilon
_{B}}-\nu _{\varepsilon -\hbar \omega _{0}+\varepsilon _{B},\varepsilon
}^{t}f_{\varepsilon }=0,~ & 0<\varepsilon <\varepsilon _{B}-\hbar \omega _{0}
\\ 
J_{ac}(f|\varepsilon )+\nu _{\varepsilon +\hbar \omega _{0}-\varepsilon
_{B},\varepsilon }^{t}f_{\varepsilon +\hbar \omega _{0}-\varepsilon
_{B}}-\nu _{\varepsilon -\hbar \omega _{0}+\varepsilon _{B},\varepsilon
}^{t}f_{\varepsilon }=0, & ~\varepsilon _{B}-\hbar \omega _{0}<\varepsilon
<\hbar \omega _{0}%
\end{array}
\label{A5}
\end{equation}%
and $\chi _{\varepsilon }$ is determined by the system 
\begin{equation}
\begin{array}{ll}
\frac{df_{\varepsilon }}{d\varepsilon }=-(\nu _{m}+\nu _{\varepsilon -\hbar
\omega _{0}+\varepsilon _{B},\varepsilon })\chi _{\varepsilon },~~ & \text{ }%
0<\varepsilon <\varepsilon _{B}-\hbar \omega _{0} \\ 
\frac{df_{\varepsilon }}{d\varepsilon }=-(\nu _{m}+\nu _{\varepsilon -\hbar
\omega _{0}+\varepsilon _{B},\varepsilon })\chi _{\varepsilon }+\widetilde{%
\nu }_{\varepsilon +\hbar \omega _{0}-\varepsilon _{B},\varepsilon }\chi
_{\varepsilon +\hbar \omega _{0}-\varepsilon _{B}},~~ & \varepsilon
_{B}-\hbar \omega _{0}<\varepsilon <\hbar \omega _{0}.%
\end{array}
\label{A6}
\end{equation}


\begin{thebibliography}{99}
\bibitem{1} R.F. Kazarinov and R.A. Suris, Sov. Phys.-Semicond. \textbf{5},
707 (1971) [Fiz. Tekh. Poluprov. \textbf{5}, 797 (1971)]; R.A. Suris and
B.S. Shchamkhalova, Sov. Phys.-Semicond. \textbf{18} 738 (1984), [Fiz. Tekh.
Poluprov. \textbf{18}, 1178 (1984)].

\bibitem{2} L.L. Bonilla and H.T. Grahn, Reports on Prog. in Phys. \textbf{68%
}, 577 (2005); R.C. Iotti and F. Rossi, Rep. Prog. Phys. \textbf{68} 2533
(2005); A. Wacker, Phys. Rep. \textbf{357}, 86 (2002).

\bibitem{3} A. Tredicucci, R. Kohler, L. Mahler, H.E. Beere, E.H. Linfield,
and D.A. Ritchie, Semicond. Sci. Technol. \textbf{20}, S222 (2005).

\bibitem{4} J. Faist, D. Hofstetter, M. Beck, T. Aellen, M. Rochat, and S.
Blaser, IEEE J. of Quant. Electr. \textbf{38}, 533 (2002); C. Gmachl, F.
Capasso, D.L. Sivco, and A.Y. Cho, Reports on Progr. in Phys. \textbf{64},
1533 (2001).

\bibitem{5} M.S. Vitiello, V. Spagnolo, G. Scamarcio, A. Lops, Q. Yang, C.
Manz, J. Wagner, Appl. Phys. Lett., \textbf{90}, 121109 (2007); M.S.
Vitiello, G. Scamarcio, V. Spagnolo, C. Worral, H.E. Beere, D.A. Ritchie, C.
Sirtori, J. Alton and S. Barbieri, Appl. Phys. Lett., {\ 89}, 131114 (2006).

\bibitem{6} C. Jirauschek, G. Scarpa, P. Lugli, M. S. Vitiello and G.
Scamarcio, J. Appl. Phys. \textbf{101}, 086109, (2007); A.Lops, V.Spagnolo,
G.Scamarcio, J. Appl. Phys. , \textbf{100}, 043109 (2006).

\bibitem{7} F.T. Vasko and A.V. Kuznetsov, \textit{Electron States and
Optical Transitions in Semiconductor Heterostructures} (Springer, New York,
1998).

\bibitem{8} V.F. Elesin and E.A. Manykin, JETP Lett. \textbf{3}, 15 (1966);
Sov. Phys. JETP \textbf{23}, 917 (1966); V.F. Elesin, Physics - Uspekhi 
\textbf{48}, 183 (2005); H. J. Stocker, Phys. Rev. Lett. \textbf{18}, 1197
(1967).

\bibitem{9} A. C. Durst and S.H. Girvin, Science \textbf{304}, 1752 (2004);
V.I. Ryzhii, Physics - Uspekhi \textbf{48}, 191 (2005); S.I. Dorozhkin,
Physics - Uspekhi \textbf{48}, 198 (2005).

\bibitem{10} F.T. Vasko, JETP Lett. \textbf{79}, 431 (2004); O.E. Raichev
and F.T. Vasko, Phys. Rev. B \textbf{73}, 075204 (2006).

\bibitem{11} R.G. Mani, J.H. Smet, K. von Klitzing, V. Narayanamurti, W.B.
Johnson, and V. Umansky, Nature \textbf{420}, 646 (2002); M.A. Zudov, R.R.
Du, L.N. Pfeiffer, and K.W. West, Phys. Rev. Lett. \textbf{90}, 046807
(2003).

\bibitem{12} F.T. Vasko and O.E. Raichev, \textit{Quantum Kinetic Theory and
Applications} (Springer, New York, 2005).

\bibitem{13} In Fig. 3, for $T_{ph}=$4.2 K, the initial distribution
function is $f_{1\varepsilon =0}\simeq $0.11 (if $T$=5 meV and $\nu _{m}=$1
ps$^{-1}$), 0.09 (if $T$=5 meV and $\nu _{m}=$0.5 ps$^{-1}$ or if $T$=3.5
meV and $\nu _{m}=$1 ps$^{-1}$), and 0.07 (if $T$=3.5 meV and $\nu _{m}=$0.5
ps$^{-1}$). For $T_{ph}=$20 K, $f_{1\varepsilon =0}\simeq $0.045 (if $T$=5
meV and $\nu _{m}=$1 ps$^{-1}$), 0.033 (if $T$=5 meV and $\nu _{m}=$0.5 ps$%
^{-1}$ or if $T$=3.5 meV and $\nu _{m}=$1 ps$^{-1}$), and 0.025 (if $T$=3.5
meV and $\nu _{m}=$0.5 ps$^{-1}$).

\bibitem{14} P. Hyldgaard and J.W. Wilkins, Phys. Rev. B \textbf{53}, 6889
(1996); M. Hartig, J.D. Ganiere, P.E. Selbmann, B. Deveaud, and L. Rota, 
\textit{ibid.} \textbf{60}, 1500 (1999); K. Kempa, P. Bakshi, J.
Engelbrecht, and Y. Zhou, \textit{ibid.} \textbf{61}, 11083 (2000).

\bibitem{15} P. Kinsler, R. M. Kelsall, and P. Harrison, Physica B \textbf{%
263}, 507 (1999).

\bibitem{16} H. Haken, \textit{Synergetics} (Springer, Berlin, 1983); E.
Scholl, \textit{Nonequilibrium Phase Transitions in Semiconductors}
(Springer, Berlin, 1987).
\end{thebibliography}
\end{document}